\documentclass{article}
\usepackage{spconf,amsmath,graphicx,hyperref}
\usepackage{amssymb}
\usepackage{multirow}
\usepackage{threeparttable}
\usepackage{booktabs}
\usepackage{cite}
\usepackage{array}
\usepackage{pifont}      
\usepackage{xcolor}

\usepackage[caption=false,font=footnotesize]{subfig} 
\graphicspath{{fig/}}

\title{Neural Network-Based Time-Frequency-bin-wise Linear Combination of Beamformers for~underdetermined~Target~Source~Extraction}
\name{
  Changda Chen$^{1}$,
  Yichen Yang$^{1,2}$,
  Wei Liu$^{1,3}$,
  and Shoji Makino$^{1}$
}
\address{
  $^{1}$Waseda University, Japan
  \\
    $^{2}$Northwestern Polytechnical University, Xi’an, China
  \\
     $^{3}$School of Electronic Information, Wuhan University, Wuhan, China
}
\begin{document}
\ninept
\maketitle
\begin{abstract}
Extracting a target source from underdetermined mixtures is challenging for beamforming approaches. Recently proposed time-frequency-bin-wise switching (TFS) and linear combination (TFLC) strategies mitigate this by combining multiple beamformers in each time-frequency (TF) bin and choosing combination weights that minimize the output power. However, making this decision independently for each TF bin can weaken temporal-spectral coherence, causing discontinuities and consequently degrading extraction performance. In this paper, we propose a novel neural network-based time-frequency-bin-wise linear combination (NN-TFLC) framework that constructs minimum power distortionless response (MPDR) beamformers without explicit noise covariance estimation. The network encodes the mixture and beamformer outputs, and predicts temporally and spectrally coherent linear combination weights via a cross-attention mechanism. On dual-microphone mixtures with multiple interferers, NN-TFLC-MPDR consistently outperforms TFS/TFLC-MPDR and achieves competitive performance with TFS/TFLC built on the minimum variance distortionless response (MVDR) beamformers that require noise priors.
\end{abstract}
\section{Introduction}
\label{sec:intro}

In acoustic signal processing, multi-microphone arrays provide spatial diversity for target source extraction (TSE) from observed signals \cite{benesty2006book, benesty2017book,makino2018book}. Classical spatial filters such as the minimum power distortionless response (MPDR) and minimum variance distortionless response (MVDR) beamformers \cite{li2003MVDR, benesty2008MVDR}, utilize array geometry to suppress interference while preserving the source signal from the target direction of arrival (DOA). Recently, neural network-based beamforming has advanced along several primary directions: estimating time-frequency (TF) masks to derive the noise covariance \cite{heymann2016neuralBFmask, erdogan2016neuralBFmask, xiao201neuralBFmask}; predicting the target relative transfer function (RTF) to strengthen MPDR or MVDR beamformers \cite{aroudi2021neuralBFrtf, yang2025neuralBFrtf}; and directly learning beamformer weights \cite{zhang2021neuralBFadl}. In parallel, time-domain methods such as FaSNet achieve learned beamforming on the waveform for lower latency \cite{luo2019neuralBFfasnet}.

However, in underdetermined situations, where the number of active sources exceeds the number of microphones, the performance of beamforming methods severely degrades. A recently proposed strategy is to exploit TF sparsity by constructing multiple candidate beamformers that share the same target steering while placing nulls at different DOAs of the interferers. Time-frequency-bin-wise switching (TFS) \cite{yamaoka2018TFS,yamaoka2019TFS} makes a hard choice, selecting one beamformer in each TF bin, whereas time-frequency-bin-wise linear combination (TFLC) \cite{yamaoka2021TFLC} uses soft weights to combine multiple beamformers. 
Both TFS and TFLC choose combination weights that minimize the output power (min-selection) in each TF bin under a distortionless constraint and can be extended to MPDR or MVDR \cite{yamaoka2019TFS}. 
Here, MVDR typically requires prior information of the interferers to make the combination directly minimize their power, which is rarely attainable in practice. MPDR removes this requirement, but minimizing observed power risks suppressing the target. Moreover, independent per-bin selection weakens TF correlation, leading to spectral discontinuities and consequently poorer extraction performance.
By contrast, neural networks can aggregate information over frequency bins within each frame, yielding more consistent weights and a better trade-off between target leakage and interference reduction. Existing neural beam combination approaches usually fix a set of pre-designed beamformers and learn a global selection at the frame level or spectrogram level \cite{chen2018combination,liu2022combination} to perform source separation, which cannot fully exploit TF sparsity and generally relies on the beamformers designed in overdetermined situations.

To this end, we propose a neural network-based time-frequency-bin-wise linear combination (NN-TFLC) framework that encodes the mixture and each beamformed signal with inplace convolutional gated linear unit (ICGLU) \cite{liu2021ICGLU,bai2025ICGLU} and frequency-independent Bi-LSTM to preserve the full TF resolution.
 A cross-attention module then consumes mixture and beam features to produce combination weights with temporal and spectral context in each TF bin. We use the prior RTF to form phase cues and to update several candidate MPDR beamformers without explicitly estimating noise covariances. The pipeline supports a variable number of input beamformers and requires only a single update step. In experiments, NN-TFLC-MPDR consistently outperforms TFS/TFLC-MPDR and achieves competitive performance with TFS/TFLC-MVDR in dual-microphone scenarios where multiple interferers coexist.

\begin{figure*}[t]
  \centering
  \subfloat[\label{fig:all-overview}]{
    \includegraphics[width=0.78\linewidth]{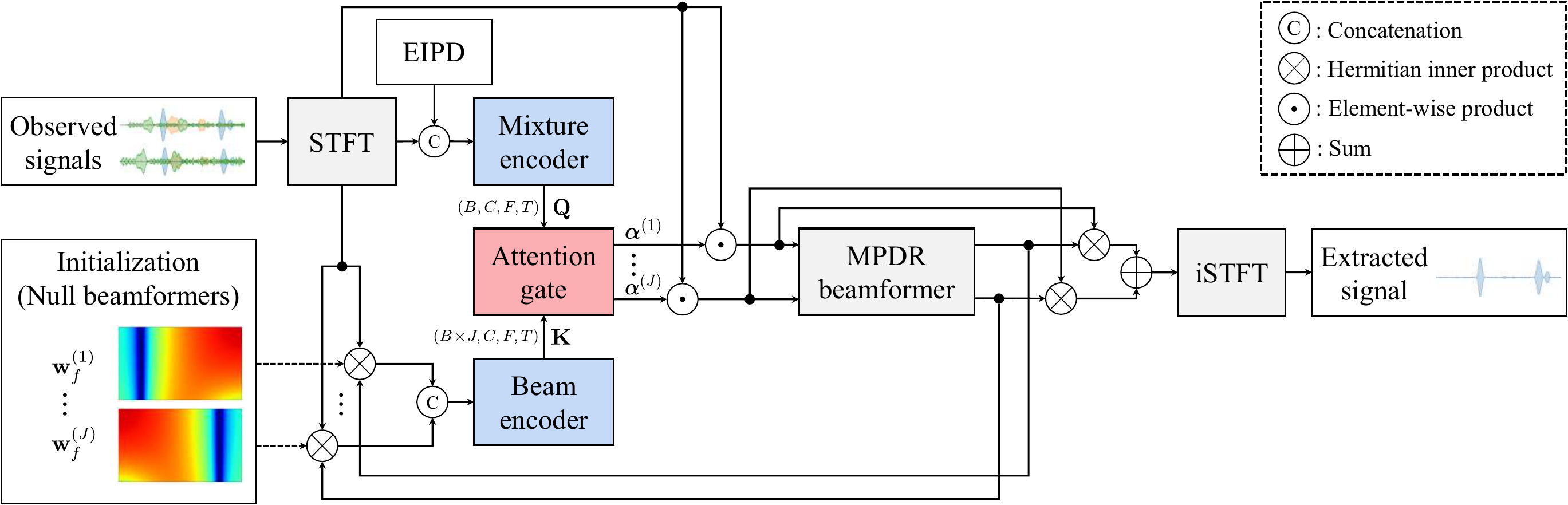}}
    \vspace{-23pt}
  \\
  \subfloat[\label{fig:all-encoder}]{
    \includegraphics[width=0.44\linewidth]{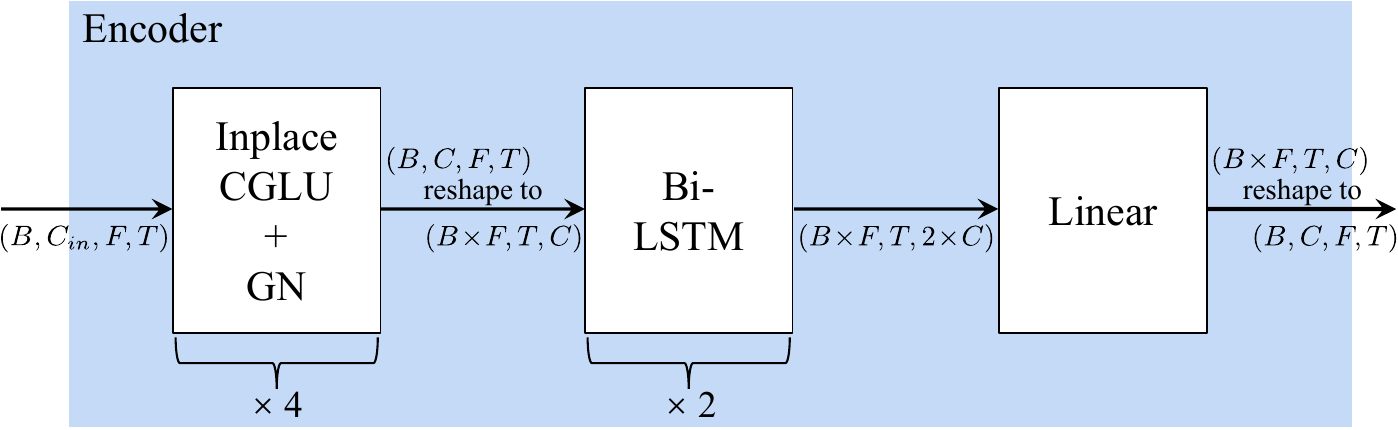}}
  \hfill
  \subfloat[\label{fig:all-att}]{
    \includegraphics[width=0.44\linewidth]{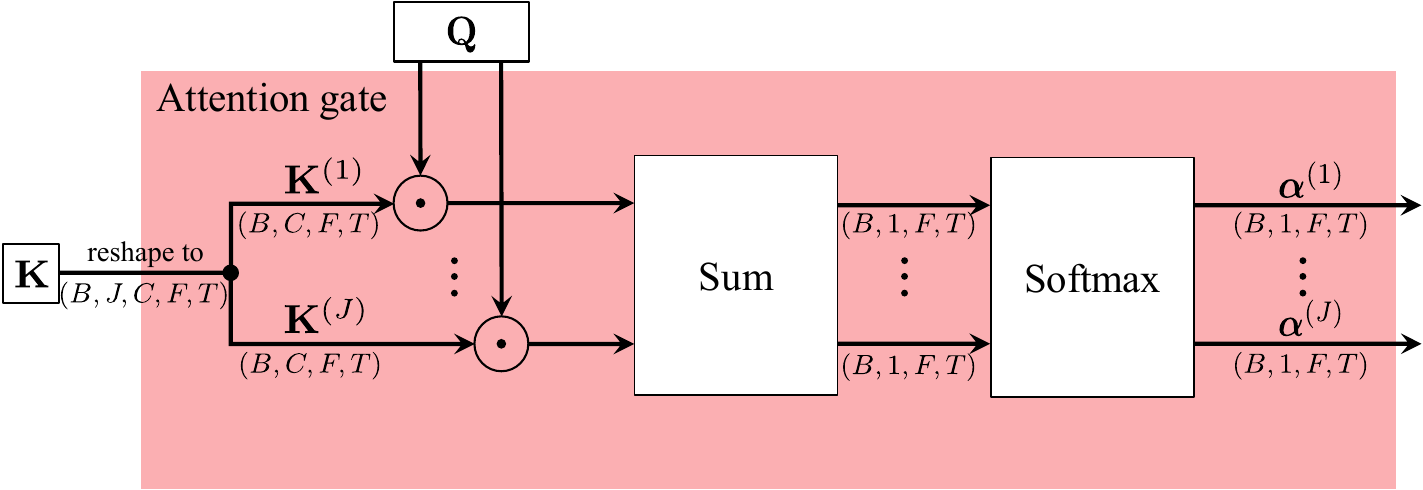}}
      \vspace{-10pt} 
  \caption{Overview and detailed components of the proposed framework.  
  (a) Overall structure of NN-TFLC-MPDR; (b) Structure of mixture encoder and beam encoder; (c) Structure of attention gate.}
  \label{fig: Model structrue}
    \vspace{-10pt} 
\end{figure*}

\section{TFS and~TFLC~of~beamformers}
\label{sec:TFS and TFLC}

Consider an $M$-element microphone array observing one target source, \mbox{$N-1$} interferers, and additive background noise.  The observation in the short-time Fourier transform (STFT) domain is
\begin{align}
\mathbf{x}_{f,t} = \mathbf{a}_{f}  S_{\text{ref},f,t} +  \sum_{n=1}^{N-1} \mathbf{h}_{n,f} I_{n,f,t} + \mathbf{n}_{f,t},
\end{align}
where $f = 1,~\ldots,~F$ and $t = 1,~\ldots,~T$ are the indices of the frequency bins and time frames, respectively, $\mathbf{x}_{f,t} \in \mathbb{C}^{M}$ contains the observed signals, $\mathbf{a}_f \in \mathbb{C}^{M }$ is the RTF of the target source signal, and $S_{\text{ref},f,t}\in \mathbb{C}$ is the target source image at the reference microphone, $I_{n,f,t}\in\mathbb{C}$ and $\mathbf{h}_{n,f}\in \mathbb{C}^{M }$ denote the signal of the $n$-th interferer and its corresponding transfer function, respectively, and $\mathbf{n}_{f,t}\in \mathbb{C}^{M}$ is the additive background noise. Our goal is to estimate $S_{\text{ref},f,t}$.

Theoretically, an \(M\)-dimensional time-invariant beamformer can impose at most \(M\) independent linear constraints (one unit response plus \(M-1\) nulls), so when more than \(M-1\) interferers are simultaneously active, conventional beamforming cannot fully suppress them. To address this limitation, we consider combining different beamformers to process the observed signals.

Under the \(M\)-disjoint orthogonality assumption mentioned in \cite{yamaoka2021TFLC}, at most \(M-1\) sources are active in each TF bin. In the dual-microphone scenarios, this assumption reduces to W-disjoint orthogonality \cite{jourjine2000W-DO,yilmaz2004W-DO}. Therefore, we construct $J$ ($\geq N-1$) candidate beamformers, each suppressing one interferer, and estimate the target via a combination in each TF bin, expressed as
\begin{align}
    \hat{S}_{\text{ref},f,t} =\sum_{j=1}^J \alpha_{f,t}^{(j)} (\mathbf{w}_f^{(j)})^\mathsf{H} \mathbf{x}_{f,t}
    \label{eq:estimated target},
\end{align}
where $\alpha_{f,t}^{(j)}$ is the corresponding linear weight of the $j$-th beamformer $\mathbf{w}_f^{(j)}$, satisfying $0\leq \alpha_{f,t}^{(j)} \leq 1$ and $\sum_{j=1}^J \alpha_{f,t}^{(j)} =1$, and $(\cdot)^\mathsf{H}$ denotes the Hermitian transpose. Estimating the linear weights in each TF bin reduces to output-power minimization: the hard case $\alpha_{f,t}^{(j)}\in\{0,1\}$ yields TFS~\cite{yamaoka2018TFS}, whereas the soft case $\alpha_{f,t}^{(j)}\in[0,1]$ yields TFLC~\cite{yamaoka2021TFLC}.

In practice, it is difficult to directly obtain predesigned beamformers whose nulls are precisely aligned with the DOAs of the interferers. Therefore,  prior work \cite{yamaoka2019TFS,yamaoka2021TFLC} adopts an iterative update strategy, where beamformers initialized with random values or with nulls placed in random DOAs are used as the starting points. Then the \mbox{linear} weights $\alpha_{f,t}^{(j)}$ are updated, and the updated weights are applied as masks to derive new beamformers, as
\begin{align}
\mathbf{w}_f^{(j)} = 
\frac{(\mathbf{\Phi}_{f}^{(j)})^{-1}\mathbf{a}_f}
{\mathbf{a}_f^\mathsf{H} \,(\mathbf{\Phi}_{f}^{(j)})^{-1}\mathbf{a}_f}
 \ \ \text{for} \ \  1\leq j \leq J
\label{eq:update MVDR},
\end{align}
where $\mathbf{\Phi}_f^{(j)} = 
\mathbb{E}[(\alpha_{f,t}^{(j)}\mathbf{x}_{f,t})
(\alpha_{f,t}^{(j)}\mathbf{x}_{f,t})^\mathsf{H}]$ is the covariance matrix for the $j$-th beamformer. Here, when \(\mathbf{x}_{f,t}\) is the full mixture including the target, this update corresponds to MPDR, and when \(\mathbf{x}_{f,t}\) contains only the noise components, this update corresponds to MVDR.

\section{Proposed method}
\label{sec:Propose}
\subsection{An overview of the proposed framework}
The overall structure of the proposed NN-TFLC-MPDR is illustrated in Fig.~\ref{fig: Model structrue}. Given dual-microphone observed signals, the waveforms are transformed to the STFT domain. The cosine and sine of the expected inter-channel phase difference (EIPD) \cite{wang2018IPD} derived from the RTF serve as the target phase cues, which are concatenated with the real and imaginary parts of the mixture along the channel dimension and fed into a mixture encoder. We prepare $J$ null beamformers $\{ \mathbf{w}_f^{(j)} \}_{j=1}^J$ with randomly selected null DOAs $\{ \theta^{(j)} \}_{j=1}^J$ to provide initial directional information. Likewise, the real and imaginary parts of their beamformed signals are concatenated along the channel dimension and encoded by a shared beam encoder. 

An attention gate takes the outputs of the mixture encoder and beam encoder as its inputs, and yields TF-bin-wise linear weights. We adopt a single iteration, in which the weights are used once to refine the MPDR beamformers. The updated candidate beamformed signals are re-encoded, and the same attention gate then yields the final combination weights, which are used to synthesize the estimated target source signal.

\subsection{Mixture encoder and Beam encoder}
Each initial null beamformer is constructed with two linear constraints: a unit response toward the target DOA, and a null toward a DOA sampled from the angular range of the interferers. In the dual-microphone scenarios, this construction is determined.
 
Both the mixture encoder and beam encoder share the same architecture in Fig.~\ref{fig:all-encoder}. The encoder begins with four ICGLU blocks \cite{liu2021ICGLU}, whose stride is set to 1 so that the TF resolution does not change. This preserves an alignment with the later prediction of TF-bin-wise linear weights. In the beam encoder, we stack $J$ beamformed signals along the batch dimension to share weights, enabling the model to naturally support a variable number of beamformer inputs. In addition, each ICGLU block is followed by a batch-size-agnostic group normalization (GN) \cite{wu2018GN}, which is more robust than batch normalization due to the distribution shifts induced by different beam patterns. After GN, the exponential linear unit (ELU) acts as the activation function.
 
 Temporal context is then modeled by a two-layer frequency-independent Bi-LSTM shared across frequency bins to consistently capture the time-delay characteristics over frequencies. A final linear layer halves the channel dimension, and the features are reshaped back to the original TF structure before entering the attention gate. 

\subsection{Attention gate}
In each TF bin, a cross-attention mechanism is employed. The linear weights of every candidate beamformer are obtained by the attention gate using softmax on a scaled dot product, expressed as 
\begin{align}
    \alpha_{f,t}^{(j)}
= \frac{\exp\!\big( \mathbf{Q}_{f,t}^\mathsf{T}\mathbf{K}_{f,t}^{(j)} /  \sqrt{C} \big)}
       {\sum_{j'=1}^{J}\exp\!\big( \mathbf{Q}_{f,t}^\mathsf{T} \mathbf{K}_{f,t}^{(j')}  / \sqrt{C} \big)} \ \ \text{for} \ \  1\leq j \leq J
       ,
\end{align}
where $\mathbf{Q}_{f,t} \in \mathbb{R}^C$ is the query from the mixture encoder, $\mathbf{K}_{f,t}^{(j)} \in \mathbb{R}^C$ is the $j$-th key from the beam encoder,  $(\cdot)^\mathsf{T}$ denotes the transpose, and $C$ is the encoded channel dimension. 

The linear weights $\alpha_{f,t}^{(j)}$ first serve as masks to form masked covariances and update $J$ MPDR beamformers via \eqref{eq:update MVDR}, where  \(\mathbf{x}_{f,t}\) includes the target. Then, the refined beamformers are fed back through the shared beam encoder, re-encoded, and passed through the attention gate to obtain the final combination weights, which yield the target estimate in \eqref{eq:estimated target}, followed by inverse STFT (iSTFT). 

\subsection{Training strategy}
We use two loss functions to train the neural network: a scale-invariant signal-to-distortion ratio (SI-SDR)\cite{le2019sdr}, and an entropy regularization term.

To help the network learn to recover the target source signals from the mixture, the negative SI-SDR loss is employed, expressed as
\begin{align}
\mathcal{L}_{\text{SI-SDR}}
= -10 \log_{10}\!\left[
\frac{\mathbb{E}(|\beta \mathbf{s}|^2)}
     {\mathbb{E}(|\beta\mathbf{s}-\hat{\mathbf{s}}|^2)}
\right],
\end{align}
where $\mathbf{s}$ and $\hat{\mathbf{s}}$ denote the time-domain reference and enhanced signals, respectively, via iSTFT from $S_{\text{ref},f,t}$ and $\hat{S}_{\text{ref},f,t}$, and $\beta=\hat{\mathbf{s}}^\mathsf{T}\mathbf{s}/{\|\mathbf{s}\|_2^2}$ is the scaling factor. In the early stage of training, we observed that the final estimated linear weights are nearly uniform, e.g., $\alpha_{f,t}^{(j)}\approx 1/J$, which severely stalls the training process. Therefore, an entropy regularization term is proposed, expressed as
\begin{align}
\mathcal{L}_{\text{Ent}}
=
- \frac{1}{FTJ}
\sum_{f=1}^F 
\sum_{t=1}^T
\sum_{j=1}^J \alpha_{f,t}^{(j)}
\ln (\alpha_{f,t}^{(j)} + \epsilon)
,
\end{align}
where $\epsilon>0$ prevents taking $\ln0$. Minimizing this term not only helps the linear weights deviate from the uniform distribution, which accelerates training and convergence, but also encourages more decisive selection, allowing the difference in covariance statistics during the update of MPDR beamformers to be amplified, thereby enhancing the complementarity among the candidate beamformers.

The final loss function is summarized as 
\begin{align}
\mathcal{L}
= \mathcal{L}_{\text{SI-SDR}} +
\lambda\mathcal{L}_{\text{Ent}}
\label{eq:final loss},
\end{align}
where $\lambda$ is the weight of the entropy regularization term.

\section{Experiments}
\label{sec:Experiments}

\subsection{Mixing conditions}

We synthesize mixtures from clean utterances of LibriSpeech~\cite{panayotov2015librispeech}.
For each mixture, a simulated room is generated with length sampled from $[6,~10]$ m, width from $[5,~8]$ m, height from $[2.5,~3.5]$ m, and reverberation time $T_{60}$ from $[0.2,~0.5]$ s.
A dual-microphone linear array with $2$ cm spacing is randomly placed within the room at a height of $1.5$ m, and its center is at least $2.5$ m from any wall.
The source-to-array distance is chosen from $[1.5,~2.0]$ m, and the source height is between $1.4$ m and $1.6$ m.
The target DOA is selected from $[80^\circ,~100^\circ]$, whereas DOAs of the interferers are limited to $[0^\circ,~65^\circ]$ and $ [115^\circ,~180^\circ]$, with at most two interferers in each range.
Clean signals are convolved with simulated RIRs using the image method \cite{allen1979image}. Each interferer is then scaled to achieve an input SIR of $[0,~5]$ dB.
Additive noise comprises simulated diffuse noise \cite{habets2008diffuse} and white Gaussian noise. The diffuse-to-white power ratio is in $[15,~25]$ dB, and the overall additive noise is adjusted to yield an SNR in $[10,~25]$ dB with respect to the reverberant target.
All the above random choices follow a uniform distribution.

In total, we create 25,000 mixtures for training, 3,000 for validation, and 3,000 for testing, each 6-second-long. The training set contains 15,000 mixtures with two interferers (2I), 5,000 with three (3I), and 5,000 with four (4I). The validation and test sets have a similar composition with 2,000 (2I), 500 (3I), and 500 (4I).

\subsection{Model configurations}
In each ICGLU block, the kernel size is set as $(5 \times 1)$, and the number of encoded channels is set to $C=32$. All audio is sampled at 16 kHz, with a 1024-sample Hanning window and a 256-sample shift. The weight of the entropy regularization term is $\lambda=0.05$. The Adam optimizer \cite{adam2014adam} is employed, together with a StepLR scheduler that reduces the learning rate by a factor of $0.8$ every $10$ epochs. The batch size is set to $B=4$. 

In our proposed method, we evaluate two model variants. \textbf{NN-TFLC-MPDR (w/o Full)} is trained only on the 2I subset with an initial learning rate of $6\times10^{-4}$, and its inputs are two beamformers whose null DOAs are randomly drawn from $[10^\circ,~55^\circ]$ and $[125^\circ,~170^\circ]$ respectively, to augment the directional diversity of the training data. The best model is chosen by the average SI-SDR on the 2I validation split over 100 epochs. \textbf{NN-TFLC-MPDR (w/ Full)} is initialized from the 2I model and then continues training on the 3I and 4I subsets, with an initial learning rate of $2\times10^{-4}$, using four input beamformers with nulls randomly drawn from $[10^\circ,~30^\circ]$, $[35^\circ,~55^\circ]$, $[125^\circ,~145^\circ]$, and $[150^\circ,~170^\circ]$. Its best model is selected by the average SI-SDR on the entire validation set over another 100 epochs.

For validation and testing, we fix the null angles. In the 2I case, the nulls of two initial beamformers are set to $32.5^\circ$ and $147.5^\circ$. In the 3I and 4I cases, we use $16.25^\circ$, $\ 48.75^\circ$, $\ 131.25^\circ$, and $\ 163.75^\circ$.

\begin{table*}[t]
  \centering
  \caption{Average SI-SDR (dB), SI-SIR (dB), and PESQ scores for processed signals with 2/3/4 interferers (mean$\scriptsize\textcolor{black!55}{\pm\text{standard deviation}}$).}
  \label{tab:performance}

  \setlength{\tabcolsep}{2.4pt}
  \renewcommand{\arraystretch}{0.94}
  \setlength{\aboverulesep}{0pt}
  \setlength{\belowrulesep}{0pt}
  \setlength{\abovetopsep}{0pt}
  \setlength{\belowbottomsep}{0pt}
  \footnotesize

  \resizebox{0.98\textwidth}{!}{%
    \setlength{\aboverulesep}{0ex}
    \setlength{\belowrulesep}{0.1ex}

    \begin{tabular}{l !{\vrule width 0.5pt}
                    c c c !{\vrule width 0.5pt}
                    c c c !{\vrule width 0.5pt}
                    c c c}
      \toprule
      \textbf{Method} &
        \multicolumn{3}{c!{\vrule width 0.5pt}}{\textbf{2I (2 beamformers)}} &
        \multicolumn{3}{c!{\vrule width 0.5pt}}{\textbf{3I (4 beamformers)}} &
        \multicolumn{3}{c}{\textbf{4I (4 beamformers)}} \\
      \cmidrule(l{0.2em}r{0.2em}){2-4}
      \cmidrule(l{0.2em}r{0.2em}){5-7}
      \cmidrule(l{0.2em}r{0.2em}){8-10}
       & \textbf{SI-SDR} & \textbf{SI-SIR} & \textbf{PESQ}
       & \textbf{SI-SDR} & \textbf{SI-SIR} & \textbf{PESQ}
       & \textbf{SI-SDR} & \textbf{SI-SIR} & \textbf{PESQ} \\
      \midrule
      Unprocessed                 & -0.81$\scriptsize\textcolor{black!55}{\pm\text{1.00}}$ & -0.69$\scriptsize\textcolor{black!55}{\pm\text{1.04}}$ & 1.13$\scriptsize\textcolor{black!55}{\pm\text{0.07}}$ & -2.48$\scriptsize\textcolor{black!55}{\pm\text{0.82}}$ & -2.40$\scriptsize\textcolor{black!55}{\pm\text{0.83}}$ & 1.09$\scriptsize\textcolor{black!55}{\pm\text{0.07}}$ & -3.88$\scriptsize\textcolor{black!55}{\pm\text{0.77}}$ & -3.81$\scriptsize\textcolor{black!55}{\pm\text{0.78}}$ & 1.09$\scriptsize\textcolor{black!55}{\pm\text{0.09}}$ \\
      \midrule
      MVDR                        &  0.93$\scriptsize\textcolor{black!55}{\pm\text{1.07}}$ &  2.61$\scriptsize\textcolor{black!55}{\pm\text{1.40}}$ & 1.16$\scriptsize\textcolor{black!55}{\pm\text{0.08}}$& -0.95$\scriptsize\textcolor{black!55}{\pm\text{0.90}}$ &  0.33$\scriptsize\textcolor{black!55}{\pm\text{1.09}}$ & 1.11$\scriptsize\textcolor{black!55}{\pm\text{0.08}}$ & -2.46$\scriptsize\textcolor{black!55}{\pm\text{0.81}}$ & -1.35$\scriptsize\textcolor{black!55}{\pm\text{0.95}}$ & 1.09$\scriptsize\textcolor{black!55}{\pm\text{0.06}}$ \\
      TFS-MVDR~\cite{yamaoka2019TFS}   &  4.16$\scriptsize\textcolor{black!55}{\pm\text{1.38}}$&  8.35$\scriptsize\textcolor{black!55}{\pm\text{2.16}}$& 1.24$\scriptsize\textcolor{black!55}{\pm\text{0.12}}$ &  3.98$\scriptsize\textcolor{black!55}{\pm\text{1.29}}$&  8.62$\scriptsize\textcolor{black!55}{\pm\text{1.81}}$& 1.22$\scriptsize\textcolor{black!55}{\pm\text{0.13}}$&  2.84$\scriptsize\textcolor{black!55}{\pm\text{1.09}}$&  6.88$\scriptsize\textcolor{black!55}{\pm\text{1.59}}$& 1.16$\scriptsize\textcolor{black!55}{\pm\text{0.08}}$\\
      TFLC-MVDR~\cite{yamaoka2021TFLC} &  4.52$\scriptsize\textcolor{black!55}{\pm\text{1.43}}$ &  8.04$\scriptsize\textcolor{black!55}{\pm\text{2.02}}$ & 1.25$\scriptsize\textcolor{black!55}{\pm\text{0.13}}$ &  4.54$\scriptsize\textcolor{black!55}{\pm\text{1.32}}$&  7.87$\scriptsize\textcolor{black!55}{\pm\text{1.61}}$& 1.23$\scriptsize\textcolor{black!55}{\pm\text{0.14}}$&  3.37$\scriptsize\textcolor{black!55}{\pm\text{1.13}}$&  6.16$\scriptsize\textcolor{black!55}{\pm\text{1.40}}$& 1.17$\scriptsize\textcolor{black!55}{\pm\text{0.08}}$\\
      \midrule
      TFS-MPDR~\cite{yamaoka2019TFS}   &  2.45$\scriptsize\textcolor{black!55}{\pm\text{1.51}}$ &  6.06$\scriptsize\textcolor{black!55}{\pm\text{2.36}}$ & 1.20$\scriptsize\textcolor{black!55}{\pm\text{0.10}}$ &  0.03$\scriptsize\textcolor{black!55}{\pm\text{1.55}}$ &  4.04$\scriptsize\textcolor{black!55}{\pm\text{2.21}}$ & 1.13$\scriptsize\textcolor{black!55}{\pm\text{0.08}}$ & -0.51$\scriptsize\textcolor{black!55}{\pm\text{1.42}}$ &  3.09$\scriptsize\textcolor{black!55}{\pm\text{1.98}}$ & 1.10$\scriptsize\textcolor{black!55}{\pm\text{0.05}}$ \\
      TFLC-MPDR~\cite{yamaoka2021TFLC} &  2.86$\scriptsize\textcolor{black!55}{\pm\text{1.55}}$ &  5.56$\scriptsize\textcolor{black!55}{\pm\text{2.12}}$ & 1.21$\scriptsize\textcolor{black!55}{\pm\text{0.10}}$ &  1.31$\scriptsize\textcolor{black!55}{\pm\text{1.58}}$&  3.82$\scriptsize\textcolor{black!55}{\pm\text{1.97}}$& 1.14$\scriptsize\textcolor{black!55}{\pm\text{0.09}}$&  0.32$\scriptsize\textcolor{black!55}{\pm\text{1.39}}$&  2.53$\scriptsize\textcolor{black!55}{\pm\text{1.72}}$& 1.11$\scriptsize\textcolor{black!55}{\pm\text{0.05}}$\\
      \midrule
    \textbf{NN-TFLC-MPDR (w/o Full)} &     {4.80$\scriptsize\textcolor{black!55}{\pm\text{1.55}}$}&    {7.70$\scriptsize\textcolor{black!55}{\pm\text{2.19}}$} &    {1.28$\scriptsize\textcolor{black!55}{\pm\text{0.12}}$} &     {3.19$\scriptsize\textcolor{black!55}{\pm\text{1.44}}$} &    {5.85$\scriptsize\textcolor{black!55}{\pm\text{1.93}}$} &    {1.20$\scriptsize\textcolor{black!55}{\pm\text{0.11}}$} &     {1.27$\scriptsize\textcolor{black!55}{\pm\text{1.31}}$} &     {3.67$\scriptsize\textcolor{black!55}{\pm\text{1.73}}$} &    {1.14$\scriptsize\textcolor{black!55}{\pm\text{0.06}}$} \\
     \textbf{NN-TFLC-MPDR (w/ Full)}  &     {4.51$\scriptsize\textcolor{black!55}{\pm\text{1.52}}$}&    {7.00$\scriptsize\textcolor{black!55}{\pm\text{2.06}}$} &    {1.26$\scriptsize\textcolor{black!55}{\pm\text{0.12}}$} &     {4.71$\scriptsize\textcolor{black!55}{\pm\text{1.54}}$} &     {9.82$\scriptsize\textcolor{black!55}{\pm\text{2.55}}$} &    {1.26$\scriptsize\textcolor{black!55}{\pm\text{0.13}}$} &     {2.65$\scriptsize\textcolor{black!55}{\pm\text{1.52}}$} &    {7.08$\scriptsize\textcolor{black!55}{\pm\text{2.42}}$} &    {1.17$\scriptsize\textcolor{black!55}{\pm\text{0.07}}$} \\
      \bottomrule
    \end{tabular}%
  }
   \vspace{-10pt}
\end{table*}

\subsection{Results and discussions}
For comparison, MVDR, TFS-MVDR\cite{yamaoka2019TFS}, and TFLC-MVDR\cite{yamaoka2021TFLC}, TFS-MPDR\cite{yamaoka2019TFS}, and TFLC-MPDR \cite{yamaoka2021TFLC} are used as the baselines. All methods, including ours, share the same RTF, estimated as the principal eigenvector of the covariance matrix of the reverberant target source signals. For MVDR-based methods, the reverberant clean signals of interferers act as additional priors, and the min-selection rule is applied directly to minimize the interference and build the noise covariance for each beamformer, which is considered to have oracle performance in the original framework. All TFS/TFLC baselines run for five iterations.
After processing, we calculate their SI-SDR, scale-invariant signal-to-interference ratio (SI-SIR) \cite{scheibler2022cal_sdr}, and perceptual evaluation of speech quality (PESQ) scores \cite{beerends2002pesq} as the objective metrics. The first microphone of the array is set as the reference microphone, and the reverberant clean source image of the target is set as the reference signal for all the above metrics.

First, we compare the baselines with the proposed NN-TFLC-MPDR in the 2I, 3I, and 4I cases of the test set, respectively. As shown in Table~\ref{tab:performance}, the performance of a single MVDR beamformer degrades sharply in underdetermined scenes. 
In contrast, our NN-TFLC-MPDR consistently outperforms TFS/TFLC MPDR across 2I/3I/4I. Although NN-TFLC-MPDR (w/o Full) is trained only on 2I with two initial beamformers as inputs, it can still yield around 1 dB higher SI-SDR than TFS/TFLC-MPDR when more interferers exist in the environment, which demonstrates the extensibility of our model to adapt to varying numbers of input beamformers. NN-TFLC-MPDR (w/ Full) undergoes the full training process, and can then have substantial improvements in both SI-SDR and SI-SIR in 3I/4I, with a minor performance loss in 2I. Importantly, our method can match or surpass TFS/TFLC-MVDR in 2I/3I and  yields a SI-SDR gap of less than 1 dB in 4I without requiring the prior information of the interferers, while achieving comparable PESQ scores.

\begin{figure}[t]
  \centering
  \includegraphics[width=1.0\columnwidth]{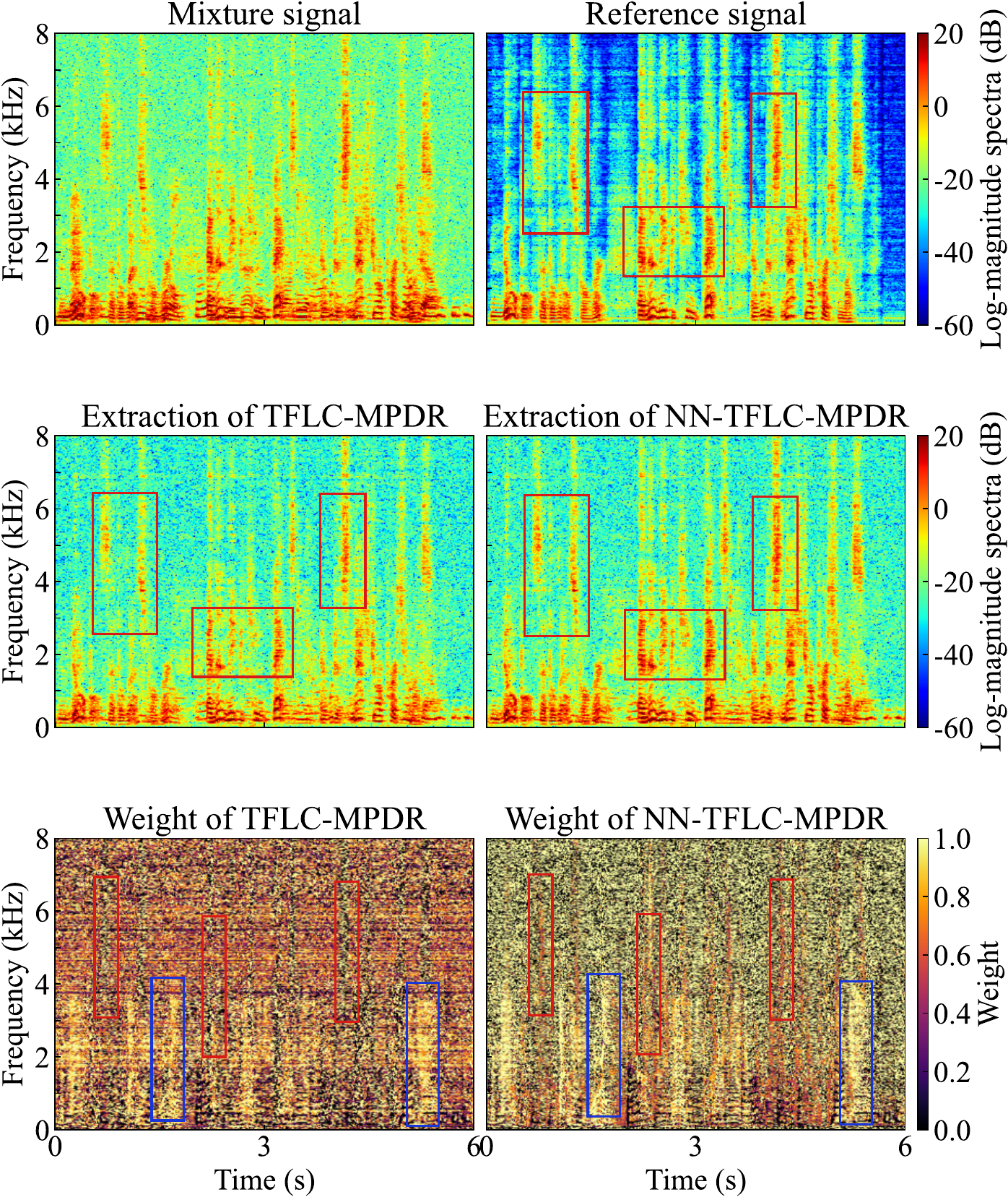}
  \caption{Log-magnitude spectrogram and combination weights of a beamformer in the TF plane for a sample from the 2I case. }
  \label{fig:Spec}
  \vspace{-10pt} 
\end{figure}

Next, we visualize the spectrograms of estimated signals and the beamformer combination weights in the TF plane processed by TFLC-MPDR and NN-TFLC-MPDR, respectively, on a sample from the 2I case. 
In Fig.~\ref{fig:Spec}, the bright bands of TFLC-MPDR appear dimmer and more fragmented than the target’s, as highlighted by the red boxes. This indicates that MPDR with per-bin min-selection does not fully preserve the target under the distortionless constraint, leading to potential suppression. While the bright bands of NN-TFLC-MPDR nearly coincide with those of the reference, which demonstrates the NN-based selection reduces target distortion and achieves a better trade-off between target preservation and suppression of interference. 

We further plot the combination weights for one of the two candidate beamformers in Fig.~\ref{fig:Spec}. It is obvious that per-bin min-selection of TFLC yields fragmented weight maps with weak cross-bin correlation. The discontinuities make the weights noisy and irregular across TF bins, potentially introducing phase inconsistencies and artifacts. By contrast, the NN-derived weights are coherent over time and frequency, and transition smoothly, especially in low-frequency regions and along several frequency bands. This behavior arises because attention in each TF bin operates on features that already encode TF context. As highlighted by the red boxes, in these target-dominant regions, the network tends to blend two beamformers rather than producing the patchwork mixing seen in TFLC-MPDR. This reduces the risk of directly selecting a beamformer whose null lies near the target, and thus helps preserve the target component. Also, in interference-dominant regions, shown by the blue boxes, the network makes more decisive selections, including at high frequencies where background noise is prominent, which strengthens beam selectivity and improves suppression of different interferers.

\section{Conclusions}
\label{sec:Conclusions}
In this paper, we present an NN-TFLC framework that constructs MPDR beamformers. It uses a neural network to estimate TF-bin-wise linear weights for combining multiple beamformers to extract the target source from underdetermined mixtures, without explicit noise covariance estimation. Across mixtures with multiple interferers, our method consistently outperforms conventional TFS/TFLC-MPDR and achieves competitive performance with TFS/TFLC-MVDR, which leverages prior information of the interferers to directly minimize their power. Moreover, experiments show that the proposed network adapts seamlessly to different numbers of input beamformers, demonstrating strong scalability and practicality.

\vfill\pagebreak

\bibliographystyle{IEEEbib}
\bibliography{refs}

\end{document}